\documentclass[11pt]{article}
\usepackage{amsmath,amssymb}
\usepackage{graphicx}
\usepackage{color}
\usepackage{mathrsfs}
\usepackage[colorlinks=true,linkcolor=blue,citecolor=blue]{hyperref}

\usepackage[margin=2.5cm]{geometry}
\usepackage{subfigure}
\usepackage{setspace}
\title{Painlev\'e solitons of AKNS system and irrational algebraic solitons of NLS equations}
\author{Man Jia$^1$, Xia-Zhi Hao$^2$, Ruo-Xia Yao$^3$ Fa-Ren Wang$^1$ and S. Y. Lou$^{1,4}$\\
\small{\it $^1$School of Physical Science and Technology, Ningbo University, Ningbo, 315211, China}\\
\small{\it $^2$School of Mathematical Sciences, Zhejiang University of Technology, Hangzhou 310014, China}\\
\small{\it $^3$School of Artificial Intelligence and Computer Science, Shaanxi Normal University, Xi'an, 710119, China} \\
\small{\it $^4$Institute of Fundamental Physics and Quantum Technology, Ningbo University, Ningbo, 315211, China}}
\date{}
\begin{document}
\maketitle

\begin{abstract}
A novel symmetry decomposition approach is introduced to derive the so-called ``Painlev\'e solitons'' of the Ablowitz-Kaup-Newell-Segur (AKNS) system. These Painlev\'e solitons propagate against a background governed by a Painlev\'e transcendent, establishing a fundamental generalization of the well-known elliptic solitons concept. We demonstrate that while elliptic solitons arise from the combination of translation invariance and square eigenfunction symmetry, a \textit{different} symmetry combination-scaling invariance, Galilean invariance, and square eigenfunction symmetry-generates ``Painlev\'e IV solitons'' for the AKNS system. This discovery represents a significant theoretical advance in integrable systems theory. By selecting special solutions of the Painlev\'e IV equation, we obtain explicit forms of several previously unknown classes of solutions for the AKNS system and the nonlinear Schr\"odinger (NLS) equation: irrational algebraic solitons, rational algebraic solitons, and parabolic cylindrical function solitons. These results dramatically expand the known solution landscape of one of the most important integrable models in mathematical physics, with broad implications for nonlinear wave phenomena across multiple physical disciplines including optics, Bose-Einstein condensates, and fluid dynamics.


\end{abstract}

\maketitle

\section{Introduction}
The study of integrable nonlinear partial differential equations (PDEs) has been a central theme in mathematical physics for over half a century. Among the most influential discoveries in this field is the inverse scattering transform (IST), a powerful method for solving certain nonlinear evolution equations. A major breakthrough in the systematic development of the IST was the formulation of the Ablowitz-Kaup-Newell-Segur (AKNS) system \cite{Ablowitz1973PRL,AKNS}. This elegant framework provides a unified approach to treat a wide class of integrable equations, including the celebrated nonlinear Schr\"odinger (NLS) \cite{NLS,NLS1}, Korteweg-de Vries (KdV) \cite{KdV,KdV1,KdV2}, modefied KdV \cite{MKdV,MKdV1,MKdV2} and sine-Gordon equations \cite{sG,sG1,sG2}, by associating them with a compatible pair of linear differential equations (a Lax pair). Within this scheme, the search for special solutions, particularly solitons, and the investigation of their interactions have revealed profound connections with other branches of mathematics, such as the theory of special functions and Painlev\'e transcendents.

The nonlinear Schr\"odinger equation,
\begin{equation}
p_t+\frac12b i p_{xx}-i|p|^2p=0,\quad i=\sqrt{-1},\ b^2=1,
\label{eq:nls}
\end{equation}
where $b=\pm 1$ corresponds to normal/abnormal dispersion, stands as a paradigmatic example of an integrable system arising in the AKNS hierarchy. The NLS equation \eqref{eq:nls} models the evolution of slowly varying wave envelopes in dispersive media and has found unparalleled applications in optics, plasma physics, and Bose-Einstein condensates. While the IST yields fundamental solitons on vanishing backgrounds, physically relevant and mathematically intriguing situations often involve solitons propagating on non-trivial, spatially or temporally varying backgrounds.

A significant step in this direction is the concept of the ``elliptic solitons'' \cite{Matveev1991}. Here, the soliton propagates on a background given by an elliptic (or cnoidal) wave, which itself is an exact periodic solution of the NLS equation. The elliptic wave is expressed in terms of Jacobi elliptic functions. The composite solution, the elliptic solitons, can be derived through sophisticated methods such as the Darboux transformation or B\"acklund transformation, applied to the finite-gap (periodic or quasi-periodic) solutions of the integrable system. Ling and Sun \cite{Ling} construct multiple elliptic solitons and analyze their asymptotic behaviors for the focusing modified KdV equation via Darboux-B\"acklund transformation. Nijhoff, Sun and Zhang \cite{NSZ-CMP-2023} introduced the concept of elliptic $N$th root of unity and established an elliptic direct linearization scheme (of Dutch version).
Elliptic $N$th root of unity can be used to defined plane wave factors for the Boussinesq equations and the equations in the Gal'fand-Dicky hierarchy.
A bilinear framework serving for elliptic solitons was established by Li and Zhang for both continuous \cite{LZ-JNS-2022} and discrete systems.
More recently, an elliptic direct linearization scheme (Fokas-Ablowitz's version) was established by Li, Sun and Zhang \cite{LSZ-N-2025} for the KP, KdV and Boussinesq equations.
The existence of such solutions is deeply tied to the underlying symmetries of the equation. Specifically, it has been shown that the combination of the space-time translation invariance and the so-called nonlocal symmetries \cite{Lou2021} such as the pseudopotential symmetries \cite{Lou1994}, residual symmetries \cite{Gao2013}, square eigenfunction symmetry \cite{Lou2012,Hu2012} is instrumental in generating these elliptic soliton solutions.

While elliptic backgrounds are associated with the first Painlev\'e transcendents in a certain asymptotic limit, a more direct and profound connection exists between integrable PDEs and the six Painlev\'e equations (PI-PVI). These second order nonlinear ordinary differential equations (ODEs), defined by their Painlev\'e property (the absence of movable critical points), play a crucial role as symmetry reductions of integrable PDEs \cite{Clarkson2006}. Furthermore, they appear in the characterization of certain special solutions. Recently, attention has turned to constructing solitons on backgrounds described by Painlev\'e transcendents. In particular, the Painlev\'e IV equation,
\begin{equation}
\frac{d^2w}{dz^2} = \frac{1}{2w}\left(\frac{dw}{dz}\right)^2 +\frac{3}{2} w^3 +4zw^2 +2(z^2 -\alpha)w + \frac{\beta}{w},
\label{eq:piv}
\end{equation}
where $\alpha$ and $\beta$ are complex parameters, emerges naturally in similarity reductions of the NLS equations, the derivative NLS equations, the Boussinesq equation \cite{Bq1872,Li2026} and other integrable systems. Its solutions, the Painlev\'e IV transcendent, provide a rich class of potential backgrounds that are asymptotically rational, algebraic, or expressible in terms of parabolic cylinder functions and error functions.

By analogy with elliptic solitons, one can define ``Painlev\'e solitons'' as solitonic excitations propagating on a background governed by a Painlev\'e transcendent. A novel symmetry decomposition approach offers a systematic way to derive such solutions \cite{Li2026}. This method leverages the interplay of multiple intrinsic symmetries of the AKNS system. Whereas elliptic solitons arise from the combination of translation invariance and the square eigenfunction symmetry, a different combination-involving \textit{scaling invariance} and \textit{Galilean invariance} alongside the square eigenfunction symmetry-leads to a different class of solutions\cite{Li2026}. For the AKNS system encompassing the NLS equation, this specific symmetry combination generates solutions known as ``Painlev\'e IV solitons.''

The Painlev\'e IV equation admits a hierarchy of special solutions for particular choices of the parameters $(\alpha, \beta)$ \cite{Clarkson2003}. By selecting these, one can obtain explicit forms of the corresponding Painlev\'e IV soliton backgrounds. For instance, when the parameters are tuned to certain integer or half-integer values, the Painlev\'e IV transcendent reduces to rational solutions or solutions expressible in terms of parabolic cylinder functions $D_{\nu}(z)$. Consequently, this yields specific types of composite wave solutions.

In this work, we introduce a fundamentally new concept: ``Painlev\'e solitons'' as solitonic excitations propagating on a background governed by a Painlev\'e transcendent. We develop a novel symmetry decomposition approach that systematically reveals how different combinations of intrinsic symmetries of the AKNS system lead to distinct classes of solutions. While elliptic solitons arise from translation invariance and square eigenfunction symmetry, we demonstrate that a different combination-involving \textit{scaling invariance} and \textit{Galilean invariance} alongside the square eigenfunction symmetry-generates ``Painlev\'e IV solitons.'' This represents a significant theoretical advance with broad implications across integrable systems theory. By selecting special solutions of the PIV equation, we obtain explicit forms of several new classes of exact solutions for the NLS equation, including previously unreported \textit{irrational algebraic solitons}, rational algebraic solitons, and parabolic cylindrical function solitons. These discoveries dramatically expand the known solution landscape of this fundamental model and provide new insights into the intricate relationship between symmetries of integrable systems, special function theory, and nonlinear wave phenomena.

\section{Symmetry Decomposition Method}
To study Painlev\'e solitons, we consider the extended AKNS system:
\begin{eqnarray}
&&p_t+\frac12b i p_{xx}-i p^2q=0,\nonumber\\
&&q_t-\frac12b i q_{xx}+i q^2p=0,\ b^2=1,\ i^2=-1, \label{eq:AKNS}
\end{eqnarray}
which is the compatibility condition of the Lax pair
\begin{eqnarray}
&&\left(\begin{array}{l} \phi \\ \psi\end{array}\right)_x=i
\left(\begin{array}{cc} - \lambda & \sqrt{b}q \\ -\sqrt{b}p & \lambda \end{array}\right)
\left(\begin{array}{l} \phi \\ \psi\end{array}\right), \label{eq:Lax}\\
&& \left(\begin{array}{l} \phi \\ \psi\end{array}\right)_t=
\left(\begin{array}{cc} - \frac12 i(2b\lambda^2+pq) & -\frac12(q_x-2i q\lambda)b^{3/2} \\ -\frac12 i(2p\lambda-i p_x)b^{3/2} & \frac12 i(2b\lambda^2+pq) \end{array}\right)
\left(\begin{array}{l} \phi \\ \psi\end{array}\right). \label{eq:Lat}
\end{eqnarray}
The NLS equation \eqref{eq:nls} is related to the AKNS system \eqref{eq:AKNS} by $q=p^*$. To find more symmetries of the AKNS system \eqref{eq:AKNS} and/or the Lax pair \eqref{eq:Lax}--\eqref{eq:Lat}, we introduce an auxiliary field $f$ with
\begin{eqnarray}
&&f_x=\phi\psi, \label{eq:fx}\\
&&f_t=-\frac12b^{3/2}(p\phi^2+q\psi^2-4b^{3/2}\lambda \phi\psi).\label{eq:ft}
\end{eqnarray}
Using the standard Lie symmetry approach \cite{Olver} for the extended AKNS system \eqref{eq:Lax}--\eqref{eq:Lat} and \eqref{eq:AKNS}, one can find the following symmetries,
\begin{equation}
\sigma\equiv \left(\begin{array}{l}
\sigma^p\\
\sigma^q\\
\sigma^{\phi}\\
\sigma^{\psi}\\
\sigma^f\\
\sigma^{\lambda}
\end{array}\right) =\left(\begin{array}{l}
P-Xp_x-Tp_t\\
Q-Xq_x-Tq_t\\
\Phi-X\phi_x-T\phi_t\\
\Psi-X\psi_x-T\psi_t\\
F-Xf_x-Tf_t\\
\delta
\end{array}\right),\label{eq:sym}
\end{equation}
where
\begin{equation}
\begin{aligned}
P&=-i b a x p+m\psi^2+\phi_0 p,\\
Q&=i b a x q-m\phi^2-2c q-\phi_0 q,\\
\Phi&=\frac12\phi(i b a x-2i\sqrt{b} m f-2c-\phi_0+n),\\
\Psi&=-\frac12\psi(i b a x+2i\sqrt{b} m f-\phi_0-n),\\
F&=-i\sqrt{b} m f^2+n f+f_0,\\
X&=a t+c x+x_0,\
T=2c t+t_0,\
\delta= -c\lambda-\frac12 ab
\end{aligned} \label{XT}
\end{equation}
with arbitrary constants $a,\ c,\ x_0,\ t_0,\ f_0,\ \lambda,\ m,\ n$ and $ \phi_0$.

A Lie point symmetry \eqref{eq:sym} with \eqref{XT} of the extended AKNS system implies that \eqref{eq:Lax}--\eqref{eq:Lat} and \eqref{eq:AKNS} are invariant under the transformation
$$\{p,\ q,\ \phi,\ \psi,\ f,\ \lambda\}\rightarrow \{p,\ q,\ \phi,\ \psi,\ f,\ \lambda\}+\epsilon \{\sigma^p,\ \sigma^q,\ \sigma^{\phi},\ \sigma^{\psi},\ \sigma^f,\ \delta\},$$
with infinitesimal parameter $\epsilon$.

The symmetry \eqref{eq:sym} includes the space translation ($x_0$ part), time translation ($t_0$ part), scaling ($c$ part), Galileo transformation ($a$ part), phase translation ($\phi_0$ part) and Mobious transformation ($m,\ n,\ f_0$ parts) invariance. Symmetry principles allow for the reduction of nonlinear mathematical physics, the discovery of special solutions (invariant solutions) and the classification of nonlinear mathematical physical equations. In essence, exploiting symmetry is not merely a computational trick; it provides a unifying framework for understanding the structure, properties, and solutions of nonlinear systems across physics and applied mathematics. In this paper, we study only the symmetry \eqref{eq:sym} invariant solutions, elliptic solitons and Painlev\'e solitons.

Combining the extended AKNS system \eqref{eq:Lax},\ \eqref{eq:Lat},\ \eqref{eq:AKNS} and the symmetry \eqref{eq:sym} with $\sigma=0$, one can establish two consistent dynamic systems, the $t$-dynamic system,
\begin{eqnarray}
f_t&=&\frac{F-\psi\phi X}T,\ \phi_1\equiv 2i\lambda(X-cx)-\phi_0,\ \phi_2\equiv\frac{\sqrt{b}}2(i n+i\phi_0-2c\lambda x),\nonumber\\
p_t&=&\frac{\sqrt{-b}X\phi p^2}{T\psi}+\frac{m\psi^2-p\phi_1}T+\frac{2b i X}{\phi\psi T^2}\big[Fp-\big(mf+b\phi_2+b^{3/2}\lambda X+\sqrt{b}\lambda^2 T\big)\psi^2\big],\nonumber\\
\phi_t&=&\frac{\sqrt{-b}Xp\phi^2}{T\psi}-\frac{i\phi}{4T^2}\big[
4bX^2+T\big(2\sqrt{b}fm+2c\lambda x-2c i+n i-\phi_0 i+6\lambda X\big)\big]+\frac{2i bFX}{T^2\psi},\nonumber\\
\psi_t&=&\frac{i}{2T}\big[2\sqrt{b}Xp\phi-\big(2\sqrt{b}fm-2c\lambda x+n i+\phi_0 i+2\lambda X\big)\psi\big],\label{eq:tdyn}
\end{eqnarray}
and the $x$-dynamic system
\begin{eqnarray}
f_x&=&\phi\psi,\nonumber\\
p_x&=&2i\lambda p-\frac{\sqrt{-b}p^2\phi}{\psi}+\frac{2i\psi}{T\phi}(\phi_2+bfm+\sqrt{b}\lambda X+b^{3/2}\lambda^2T)-\frac{2b i Fp}{T\psi\phi},\nonumber\\
\phi_x&=&-\frac{\sqrt{-b}p\phi^2}{\psi}+\frac{i\phi}{T}(2bcx+2c\lambda t+2bx_0+3\lambda t_0)-\frac{2b i}{T\psi}(nf+f_0-\sqrt{-b}mf^2),\nonumber\\
\psi_x&=&i(-\sqrt{b}p\phi+\lambda\psi),\label{eq:xdyn}
\end{eqnarray}
while the field $q$ is related to other fields by
\begin{equation}
q=-\frac{\phi^2}{\psi^2}p+\frac{2b^{3/2}(2\lambda T+bX)}{T\psi}-\frac{2\sqrt{b}F}{T\psi^2}. \label{eq:q}
\end{equation}

The exact solutions of the consistent $t$-dynamic system \eqref{eq:tdyn} and the $x$-dynamic system \eqref{eq:xdyn} can be classified to two important cases.

\subsection{Case 1: Elliptic Solitons ($c=0,\ t_0=-1$)}
In this case, the general solutions of \eqref{eq:tdyn} and \eqref{eq:xdyn} read
\begin{eqnarray}
q&=&\frac{bm}{x_0A^2H}\left[\frac12bH_{\xi}-AbH^2\tanh(\zeta)+i(2b\lambda-x_0)+i x_0\right]\exp\left(\phi_0 t+K\right),\nonumber\\
p&=&\frac{A^2x_0}{mH}\left[\frac12bH_{\xi}-bA H^2\tanh(\zeta)+i(x_0-2b\lambda)H-i x_0\right]\exp\left(-\phi_0 t-K\right),\nonumber\\
\phi&=&\sqrt{F_{1\xi}}\ \mbox{sech}(\zeta)\exp\left(\frac{\phi_0 t+K}2\right),\nonumber\\
\psi&=&\frac{x_0A^2}{m\sqrt{-b}}\sqrt{F_{1\xi}}\ \mbox{sech}(\zeta)\exp\left(-\frac{\phi_0 t+K}2\right),\nonumber\\
f&=&\frac{x_0A}{2\sqrt{-b}m}\big\{2\tanh(\zeta)+n\big\},\label{eq:f}
\end{eqnarray}
where $\zeta=F_1-x_0t,\ \xi=x+x_0t,\ A^2=\frac{n^2+4m f_0\sqrt{-b}}{4x_0^2}$ while the group invariant functions $F_1\equiv F_1(\xi),\ K\equiv K(\xi),\ H\equiv H(\xi) $ satisfy the following reduction equations
\begin{eqnarray}
&&H=F_{1\xi},\ K_{\xi}=2i(\lambda - b x_0)+\frac{2i b x_0}{H},\nonumber\\
&&H_{\xi}^2=4A^2H^4+c_1 H^3+4\big(6b x_0\lambda+i b\phi_0-6\lambda^2-x_0^2\big)H^2+8x_0(x_0-2b\lambda)H-4x_0^2.
\label{eq:Hx}
\end{eqnarray}

\subsection{Case 2: Painlev\'e IV Solitons ($c=1,\ x_0=t_0=0$)}
In this case, the solutions of the consistent $t$- and $x$-dynamic systems read ($\varphi=2i b\lambda^2 t-2i\sqrt{t}\lambda\eta+2K,\ \eta=\frac{x+2b\lambda t}{\sqrt{t}}$),
\begin{eqnarray}
p&=&-\frac{bA^2t^{\frac{\phi_0}2}}{mF_{2\eta}}\left\{2AF_{2\eta}^2\tanh\big[A(\ln(t)+F_2)\big]
+i b\eta F_{2\eta}-2i b-F_{2\eta\eta}\right\}\exp(-\varphi),\nonumber\\
q&=&-\frac{m t^{-1-\frac{\phi_0}2}}{4A^2F_{2\eta}}\left\{2AF_{2\eta}^2\tanh\big[A(\ln(t)+F_2)\big]
-i b\eta F_{2\eta}+2i b-F_{2\eta\eta}\right\}\exp(\varphi),\nonumber\\
\phi&=&\sqrt{F_{2\eta}}\ t^{-\frac12-\frac{\phi_0}4}\mbox{sech}[A(\ln(t)+F_2)]\exp\left(\frac12\varphi\right),\nonumber\\
\psi&=&\frac{2A^2}{m\sqrt{-b}}\sqrt{F_{2\eta}}\ t^{-\frac{\phi_0}4}\mbox{sech}[A(\ln(t)+F_2)]\exp\left(-\frac12\varphi\right),\nonumber\\
f&=&\frac1{2m\sqrt{-b}}\left\{n+4A\tanh[A(\ln(t)+F_2)]\right\}, \label{eq:PIV}
\end{eqnarray}
where the group invariant functions $F_2=F_2(\eta)$ and $K$ are determined by
\begin{eqnarray}
K_\eta&=&\sqrt{-b}\left(\frac{\eta}2-F_{2\eta}^{-1}\right),\nonumber\\
F_{2\eta}F_{2\eta\eta\eta}&=&2A^2F_{2\eta}^4-\frac12 i(i\eta^2-2b-2b\phi_0)F_{2\eta}^2
-4\eta F_{2\eta}+\frac32F_{2\eta\eta}^2+6.
 \label{eq:F}
\end{eqnarray}
\textbf{Remark.} It should be emphasized that all results obtained so far are valid for the AKNS system for both $b=1$ and $b=-1$. However, when imposing the constraint $p=q^*$ on all formulae to derive results for the NLS equation \eqref{eq:nls}, one must exercise care with $\sqrt{b}$ and/or $\sqrt{-b}$. Several special examples that hold for the AKNS system but not for the NLS equation will be pointed out in the next section.

\section{Explicit Solutions and Results}
\subsection{Elliptic Soliton Solutions}
From the $H$ equation \eqref{eq:Hx}, we know that $H$ is determined by a simple elliptic integral. Thus, the solution \eqref{eq:f} represents an explicit elliptic soliton solution of the AKNS system \eqref{eq:AKNS} and the NLS equation \eqref{eq:nls} for $A^4x_0^2-bm^2=0$, imaginary $\{\phi_0, \ K\}$ and real $\{H,\ F,\ \lambda,\ x_0,\ A,\ c_1,\ m,\ n\}$. The elliptic solution of the $H$ equation \eqref{eq:Hx} can be explicitly written in the form
\begin{eqnarray}
H=\frac{a_1 J+a_0}{a_3 J+a_2},
\label{eq:rH}
\end{eqnarray}
where the function $J$ may be chosen as the Weierstrass function $\mathscr{P}(d\xi,\ g_2,\ g_3)$ as well as elliptic Jacobi functions such as $\mbox{cn}^2(d\xi,\ k)$, $\mbox{sn}(d\xi,\ k)$, $\mbox{cn}(d\xi,\ k)$, $\mbox{dn}(d\xi,\ k)$ and so on for the AKNS system\cite{LouJMP1989,LouPLA1990}. However, for the NLS equation, neither $\mathscr{P}(d\xi,\ g_2,\ g_3)$ nor $\mbox{sn}(d\xi,\ k)$ is a permissible choice for $J$.

\begin{figure}[htbp]
\centering
\includegraphics[width=0.45\linewidth]{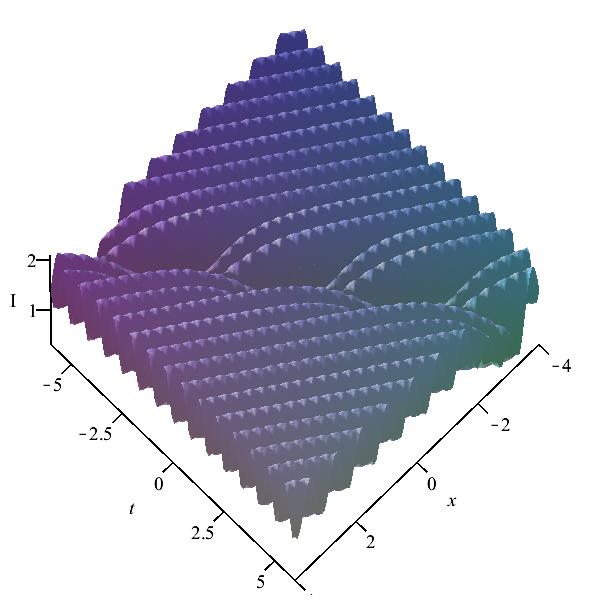}
\caption{Structure of the elliptic soliton solution \eqref{eq:f} with \eqref{eq:rcn} for the quantity $I=|p|^2=pq$.}
\label{fig:elliptic}
\end{figure}

For instance, if we write the constants $\phi_0, \ c_1,\ A,\ \lambda$ and $k$ in the form,
\begin{eqnarray}
\phi_0&=&i b\left(\frac{x_0^2}2-\frac{3a_0^2+3a_1^2+2a_0a_1}{8a_0^2a_1^2(a_0+a_1)^2}x_0^2\Delta^2
+\frac{a_0-a_1}{a_0+a_1}d^2-\frac{3a_0^2a_1^2}{8x_0^2\Delta^2}d^4\right),\nonumber\\
c_1&=&\frac{4d^2\Gamma}{\Delta(a_0+a_1)}+\frac{4x_0^2(a_2+a_3)}{a_0a_1(a_0+a_1)}
\big[\Gamma+a_2a_3(a_0+a_1)\big],\ \Gamma\equiv 2a_2a_3(a_0+a_1)+a_0a_3^2+a_1a_2^2,\nonumber\\
A^2&=&-\frac{a_2a_3(a_2+a_3)d^2}{(a_0+a_1)\Delta}-\frac{a_2a_3x_0^2(a_2+a_3)^2}{a_0a_1(a_0+a_1)^2},\nonumber\\
\lambda&=&\frac{b x_0}{2a_0}(a_0-2a_2)-\frac{b a_0a_1d^2}{4x_0\Delta}
-\frac{b x_0(a_0+3a_1)\Delta}{a_0a_1(a_0+a_1)},\nonumber\\
k^2&=&\frac{a_1}{a_0+a_1}-\frac{x_0^2\Delta^2}{a_0a_1d^2(a_0+a_1)^2},\ \Delta\equiv a_0a_3-a_1a_2,
\label{eq:rc}
\end{eqnarray}
then the elliptic periodic wave \eqref{eq:rH} becomes
\begin{eqnarray}
H=\frac{a_1 \mbox{cn}^2(d\xi,\ k)+a_0}{a_3 \mbox{cn}^2(d\xi,\ k)+a_2}
\label{eq:rcn}
\end{eqnarray}
with arbitrary constants $a_0,\ a_1,\ a_2,\ a_3,\ x_0$ and $d$. Fig. \ref{fig:elliptic} denotes the three dimensional structure of the strength $I\equiv |p|^2=pq$ for the elliptic soliton with the parameter selections, $b=d=x_0=a_3=-a_0=1,\ a_2=a_1=2,\ \xi_0=0$.

\subsection{Painlev\'e IV Solitons and Irrational Algebraic Solitons}
\begin{figure}[htbp]
\centering
\includegraphics[width=0.45\linewidth]{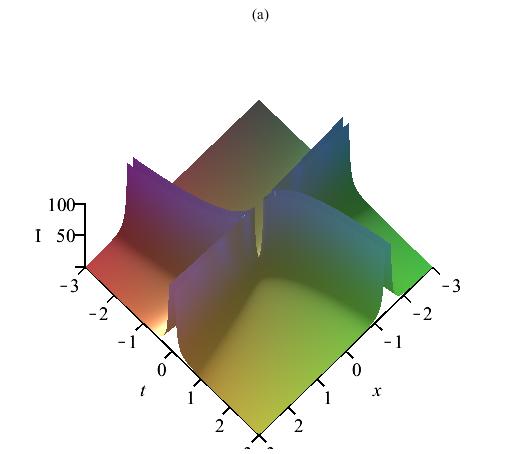}
\includegraphics[width=0.45\linewidth]{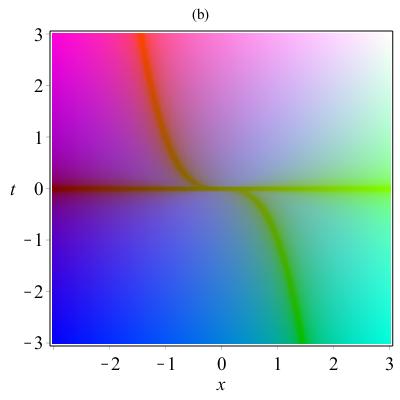}
\caption{Structure of the special Painlev\'e IV soliton, the irrational algebraic soliton related to \eqref{pPIV} with the parameter selections $b=a=1,\ \lambda=0$: (a) The three dimensional structure for the quantities $I=|p|^2=pq$ and (b) the density plot for the quantity $1/(1+I^{1/100})$.}
\label{fig:piv1}
\end{figure}

The $F$-equation \eqref{eq:F} is related to the Painlev\'e IV equation \eqref{eq:piv} by
\begin{equation}
F_{2\eta}= -2(i\delta_2+\delta_1)w^{-1},\ z=\frac12(i\delta_2+\delta_1)\eta,\ \delta_1^2=\delta_2^2=1,\ \alpha=b(1+\phi_0)\delta_2\delta_1,\ \beta=-32A^2. \label{Fw}
\end{equation}
For some special $\alpha$ and $\beta$, solutions of the Painlev\'e IV equation can be expressed by some types algebraic rational solutions, the special Whittaker functions and/or the parabolic cylinder function ${\cal{D}}_{\nu}(\zeta)$ which satisfies
\begin{equation}
u_{\zeta\zeta}=\left(\frac14\zeta^2-\nu-\frac12\right)u.\label{cylinder}
\end{equation}
One of the simple rational solution is $w=z^{-1}$ which yields $F=6\ln(a\eta),\ A=\frac1{12},\ \phi_0=-1,\ K(\eta)=\frac16 i b\eta^2+i\varphi_0$. Correspondingly, the solutions of the NLS equation reads,
\begin{equation}
p=\frac{i(3i b a t-2b\lambda t^{4/3}-4ab\lambda xt-4a\lambda^2t^2-t^{1/3}x-ax^2)}{216mt\big(ax+2ba\lambda t+t^{1/3}\big)}\ e^{\frac{i}{3t}(2b\lambda^2t^2-bx^2+2\lambda xt+6\varphi_0 t)} \label{pPIV}
\end{equation}
with the condition $\{q=p^*,\ m=\frac{\sqrt{b}}{72}\}$. The rational solutions (rogue waves) of the NLS type equations have been studied by various researchers. Different from the rational algebraic solutions, the rogue waves, of the NLS equation, the special Painlev\'e IV soliton solution \eqref{pPIV} is an irrational algebraic solution. Fig. \ref{fig:piv1} displays the three dimensional structure of the strength $I \equiv |p|^2= pq$ for the irrational algebraic soliton with the parameter selections, $b=\lambda=1$.

Two more irrational algebraic solitons of the NLS equation read,
\begin{eqnarray}
p&=&\frac{b^{3/2}(3t-i x^2)^{-b}(3t+i x^2)^b}{t(x^4+9t^2)[ax(x^4+45t^2)+45t^{5/3}]}\left\{i x(9t^2-x^4)\big[ax(x^4+45t^2)+45t^{5/3}\big]\right.\nonumber\\
&&\left.-9b a t(x^8+135t^4)+270b x^3 t^{8/3}\right\}\exp\left[\frac{-i}{3t}(b x^2+6\varphi_0 t)\right] \label{p1PIV}
\end{eqnarray}
and
\begin{eqnarray}
p&=&\frac{\left[\mu-12(b i x^2+3t)(x^4+9t^2)t\right]\left[x\mu(567x t^{7/3}-a\nu)+243b(128a t^5 x^7-21\gamma t^{10/3})\right]}{3\sqrt{b}t\alpha^2(a\beta+567t^{7/3})}e^{i\theta}, \label{p2PIV}
\end{eqnarray}
where $\alpha=x^8+54t^2x^4+81t^4,\ \beta=567t^4-126t^2x^4-x^8,\ \gamma=27t^4-18t^2x^4-x^8,\ \mu=x^8+18t^2x^4+405t^4$, $\nu=i\beta-12b t x^2(x^4+63t^2)$ and $\theta=\frac{-b}{3t}(x^2+6\varphi_0 t)$.

\begin{figure}[htbp]
\centering
\includegraphics[width=0.45\linewidth]{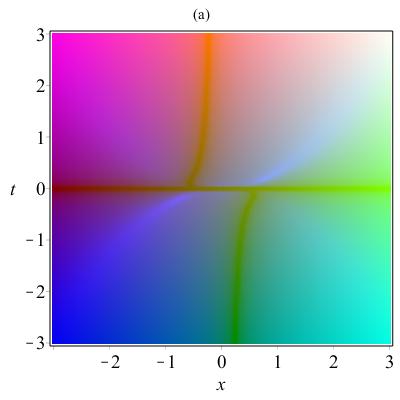}
\includegraphics[width=0.45\linewidth]{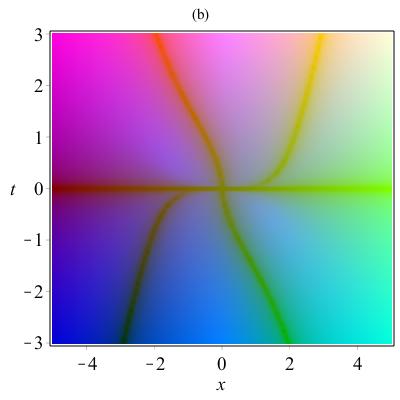}
\caption{Structures of two more special Painlev\'e IV (irrational algebraic) solitons: (a) The density plot of \eqref{p1PIV} with the parameter selections $b=1,\ a=3$ for the quantity $1/(1+|p|^{1/50})$ and (b) the density plot of \eqref{p2PIV} with the parameter selections $b=1,\ a=0.5$ for the quantity $1/(1+|p|^{1/50})$.}
\label{fig:piv23}
\end{figure}

From the Painlev\'e soliton solution \eqref{eq:PIV}, we know that for the NLS equation \eqref{eq:nls}, the condition $q=p^*$ can be satisfied only for the real function $F(\eta)$. In additional to the irrational algebraic soliton solutions, there are also some types of special Painlev\'e IV solitons, the rational algebraic solitons. For instance, it is not difficult to verify that the $F(\eta)$ equation possesses the real solution
\begin{equation}
F_2(\eta)=\frac23\ln\left(\frac{a}{\eta}-\frac{a}3\eta^3\right),\ \phi_0=-1,\ A=\frac34.\label{ratF}
\end{equation}
The rational solution of the NLS equation related to \eqref{ratF} reads
\begin{eqnarray}
p=q^*=\frac{6i a x t+2abx^3+3b}{\sqrt{b}(a x^4-3a t^2-3x)}e^{2i\varphi_0},\ m=\frac98\sqrt{b}.\label{ratp}
\end{eqnarray}

Because the $F_2$-equation \eqref{eq:F} and the Painlev\'e IV equation \eqref{eq:piv} are related via \eqref{Fw}, the argument of the Painlev\'e transcendent becomes complex with respect to the real space-time variables. Consequently, if we start directly from the solutions of the Painlev\'e IV equation, we can readily obtain solutions of the AKNS system, but not those of the NLS equation, since the complex conjugate condition $q = p^*$ is difficult to satisfy.

It is not difficult to verified that the Painlev\'e IV equation \eqref{eq:piv} possesses a special solution
\begin{eqnarray}
w=\frac{(2\nu+1){\cal{D}}_{\nu+1}(\sqrt{2}z)}{\sqrt{2}{\cal{D}}_{\nu}(\sqrt{2}z)},\ \alpha=\left(\nu-\frac12\right)\delta,\ \beta=-\frac12(2\nu+1)^2,\ \delta^2=1, \label{Dnu}
\end{eqnarray}
where ${\cal{D}}_{\nu}(\zeta)$ is the parabolic cylinder function which is a solution of \eqref{cylinder}.

Correspondingly, the $F_2$-function takes the form (with $\delta^2 = \delta_1^2 = \delta_2^2 = \delta_3^2 = 1$)
\begin{equation}\label{DF}
F_2 = \frac{4\delta}{2\nu + 1} \ln\left[ \mathcal{D}_{\nu+1} \left( \frac{\sqrt{2}\,\delta(\delta_1 + \mathrm{i}\delta_2)}{2}\,\eta \right) \right], \quad
A = \frac{\delta_3}{8}(2\nu + 1), \quad
\phi_0 = \frac{1}{2} b \delta \delta_1 \delta_2 (2\nu - 1) - 1,
\end{equation}
and the corresponding solution of the AKNS system is given by \eqref{eq:PIV} together with \eqref{DF}.
Figure 4 displays the parabolic cylinder soliton of the AKNS system for the quantity $I \equiv pq$, where $p$ and $q$ are given by \eqref{eq:PIV} and $F_2$ is specified as in \eqref{DF} with the parameter choices $\delta = \delta_1 = \delta_2 = \delta_3 = -b = 1$, $\lambda = 0$, and $\nu = 2$.

As shown in Fig. 4(b), the imaginary part of $pq$ is nonzero. Therefore, the corresponding solution belongs only to the AKNS system and does not satisfy the NLS equation.

\begin{figure}[htbp]
\centering
\includegraphics[width=0.45\linewidth]{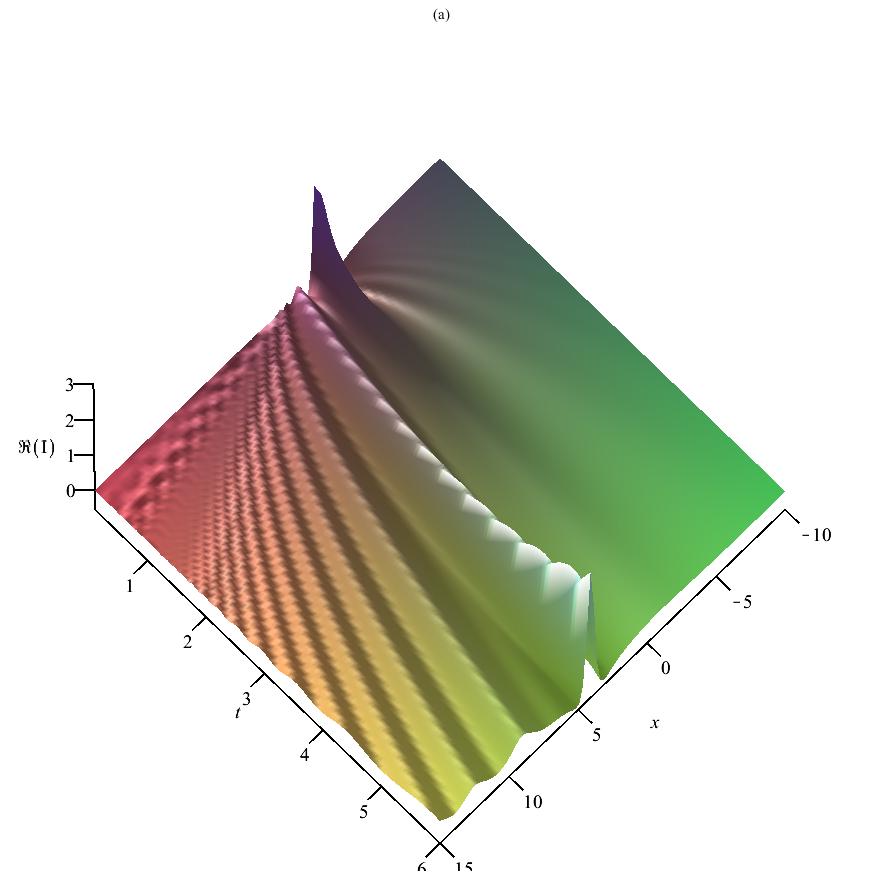}
\includegraphics[width=0.45\linewidth]{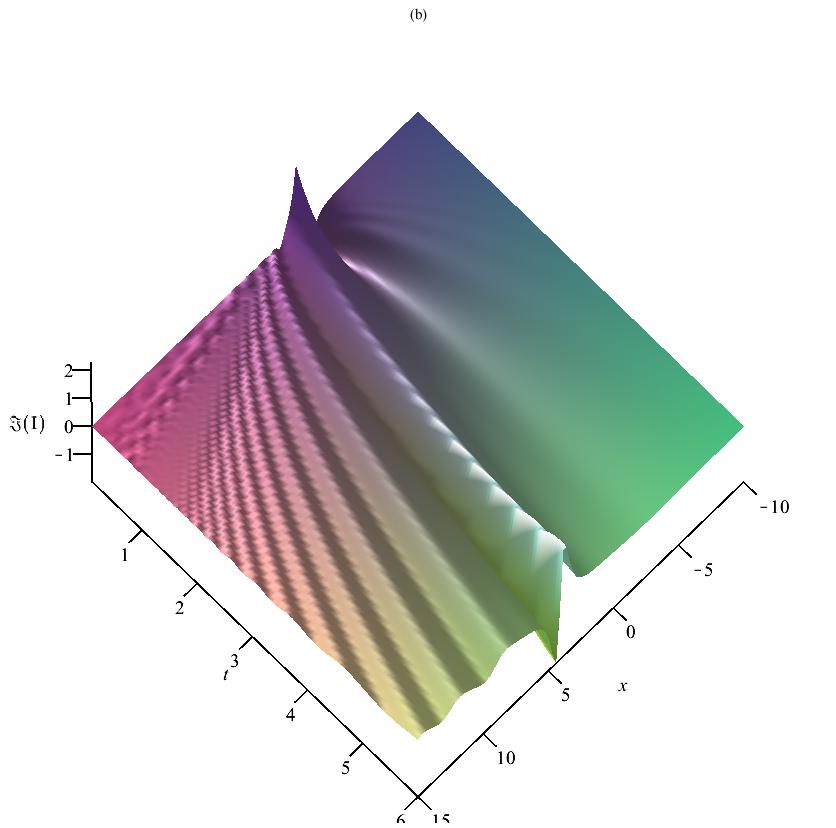}
\caption{Structure of the parabolic cylinder soliton of the AKNS system for the quantity $I \equiv pq$, where $p$ and $q$ are given by \eqref{eq:PIV} and $F_2$ is specified as in \eqref{DF} with the parameter choices $\delta = \delta_1 = \delta_2 = \delta_3 = -b = 1$, $\lambda = 0$, and $\nu = 2$: (a) Real part and (b) imaginary part of $I$.}
\label{fig:piv4}
\end{figure}

\section{Discussion and Conclusion}
In this work, we have introduced and systematically developed the fundamentally new concept of ``Painlev\'e solitons'' for the AKNS integrable hierarchy. Our approach represents a significant theoretical advance in soliton theory and integrable systems.

 We have developed a novel symmetry decomposition approach that reveals how different combinations of intrinsic symmetries of the AKNS system lead to distinct classes of solutions. While the well-established elliptic solitons \cite{Matveev1991,Ling,NSZ-CMP-2023} arise from translation invariance and square eigenfunction symmetry \cite{Lou1994,Lou1996,Lou1996a}, we have discovered that a fundamentally different symmetry combination-scaling invariance, Galilean invariance, and square eigenfunction symmetry-generates Painlev\'e IV solitons. This represents a major theoretical insight into the structure of integrable systems.

 The explicit construction of Painlev\'e solitons provides exact models for solitary waves interacting with dynamically evolving, non-periodic backgrounds governed by Painlev\'e transcendents \cite{Clarkson2006, Clarkson2003}. These solutions significantly expand the known solution landscape of the NLS equation, one of the most important integrable models in mathematical physics. The discovery of \textit{irrational algebraic solitons} is particularly noteworthy, as these represent a previously unknown class of solutions distinct from conventional rational rogue waves \cite{Kudryashov2004,Chen2012}.

 While focused on the NLS equation within the AKNS hierarchy, our symmetry decomposition method has wide applicability across integrable systems theory. The approach provides a systematic framework for generating exact solutions on Painlev\'e backgrounds, which could be extended to other Painlev\'e equations (PI-PVI) and their associated integrable systems. These results have implications for multiple physical disciplines including nonlinear optics (where the NLS describes pulse propagation in fibers), Bose-Einstein condensates (where it models matter waves), and fluid dynamics (where it describes water waves).

 Our work establishes profound connections between three fundamental areas of mathematical physics: (1) integrable systems and their symmetry structures, (2) Painlev\'e transcendents and special function theory \cite{Clarkson2003}, and (3) soliton theory and nonlinear wave phenomena. The explicit connection between specific symmetry combinations and solution types provides new insights into the algebraic-geometric structure of integrable hierarchies.

 This work opens several promising research avenues, including: investigation of the stability and interaction properties of Painlev\'e solitons; extension to other integrable systems within the AKNS hierarchy; exploration of physical applications in nonlinear optics and Bose-Einstein condensates; connection to modern topics such as non-Hermitian physics and parity-time symmetric systems; and numerical verification and experimental realization of these novel solution types.

In conclusion, we have developed a powerful symmetry-based approach to construct Painlev\'e solitons for the AKNS system, leading to the discovery of several new classes of exact solutions for the NLS equation. These results significantly expand the known solution landscape of one of the most important integrable models in mathematical physics and establish new connections between symmetry analysis, special function theory, and nonlinear wave phenomena. Our work provides both a theoretical framework and explicit solutions that will facilitate further research in integrable systems and their applications across physics.

\section*{Acknowledgments}
This work was sponsored by the National Natural Science Foundation of China (Grants No. 12235007, 12275144, 12271324, 12301315 and No. 12375003). The authors thank Professors Q. P. Liu, X. B. Hu, B. F. Feng, Y. R. Xia and Dr. Y. Li for their helpful discussions.


\end{document}